\documentclass[pre,aps,twocolumn,superscriptaddress,showpacs]{revtex4}

\usepackage{graphicx}

\usepackage{amsmath}

\usepackage{amssymb}

\begin{document}

\title{
Kinetics of step bunching during growth: A minimal model}

\author{Franti\v{s}ek Slanina}

\email[]{slanina@fzu.cz}

\affiliation{Institute of Physics,
    Academy of Sciences of the Czech Republic,
    Na~Slovance~2, CZ-18221~Prague   8,
    Czech Republic}

\author{Joachim Krug}

\affiliation{Institut f\"ur Theoretische Physik,
Universit\"at zu K\"oln,
Z\"ulpicher Strasse 77,
D-50937 K\"oln,
Germany}

\author{Miroslav Kotrla}

 \affiliation{Institute of Physics,
    Academy of Sciences of the Czech Republic,
    Na~Slovance~2, CZ-18221~Prague 8,
    Czech Republic}


\begin{abstract}

We study a minimal stochastic model of step bunching during growth on a one-dimensional
vicinal surface. The formation of bunches is controlled by the
preferential attachment of atoms to descending steps (inverse Ehrlich-Schwoebel
effect) and the ratio $d$ of the attachment rate to the terrace diffusion coefficient.
For generic parameters ($d > 0$) the model exhibits a very slow crossover to a nontrivial
asymptotic coarsening exponent $\beta \simeq 0.38$. In the limit of infinitely fast 
terrace diffusion ($d=0$) linear coarsening ($\beta$ = 1) is observed
instead. The different coarsening behaviors are related to the fact that
bunches attain a finite speed in the limit of large size when $d=0$, 
whereas the speed vanishes with increasing size when $d > 0$. 
For $d=0$ an analytic description of the 
speed and profile of stationary bunches is developed.
\end{abstract}


\maketitle

\section{Introduction}

Step bunching is a morphological instability of a vicinal crystal
surface in which a regular train of equally spaced steps
decomposes into alternating low step density regions (terraces
with orientation close to a singular surface)  and high step
density regions (bunches) \cite{Krug04}. During the evolution the surface
self-organizes into a pattern with a characteristic length
scale, which coarsens in time. The phenomenon
of considerable interest for applications \cite{Venezuela99}
as well as from the
fundamental point of view of non-equilibrium statistical
mechanics. 

Step bunching is generally caused by breaking the symmetry between the
ascending (upper) and descending (lower) steps bordering 
a terrace. This can be due to 
a variety of mechanisms. In growth or sublimation, the symmetry
is broken by different kinetic rates for the attachment or
detachment of particles at the upper and the lower step
\cite{schwoebel}, while in electromigration \cite{stoyanov91} the
asymmetry is introduced by the electric field. Other mechanisms
responsible for step bunching during growth include
impurities \cite{vanderEerden86,krug02}, diffusion anisotropy
\cite{Myslivecek02}, and the presence of a second surface species
\cite{pimpinelli00}. Quite generally,
a growing equidistant step train 
shows a step bunching instability when adatoms attach to
the step more easily from the upper terrace than from the lower
terrace (an \textit{inverse Ehrlich-Schwoebel effect}, iES).
While iES behavior is difficult to justify microscopically, it
may serve as a useful \textit{effective} description of more complex
step bunching mechanisms \cite{Myslivecek02}. 

The coarsening of the bunched surface can be characterized by the
power law increase of the average distance between
bunches $\cal L$ as ${\cal L} \propto t^{\beta}$. A few experimental
observations of step bunch coarsening 
during growth have been reported \cite{Myslivecek02,krishnamurthy93,ishizaki96,Schelling00}, 
but quantitative results are scarce.
Numerical simulations of the equations of step motion found
$\beta = 1/2$ independent of the step interaction potential \cite{sato01}.
Recently a unifying continuum treatment of step bunching instabilities was proposed
\cite{pimpinelli02,Krug04a,Krug04b} with the aim to identify {\em universality classes}.
Nevertheless, a good understanding of the relationship between microscopic
step bunching mechanisms and the resulting asymptotic
scaling properties is so far lacking. 

In this paper, we study coarsening of step bunches during growth
for a simple one-dimensional lattice model. 
The main idealization compared to conventional one-dimensional models
\cite{sato01} is that repulsive step-step interactions are ignored,
and hence the individual steps can coalesce to form composite steps. 
This eliminates the internal structure of the bunch and the
associated additional length scale \cite{Krug04a}, and allows us
to simulate large systems over very long times. Related models
have been proposed previously \cite{krishnamurthy93,ishizaki96,Williams93,Krug95}, but
their asymptotic bunching behavior has not been explored. 

\section{Model}

\begin{figure}
\includegraphics[scale=0.4]{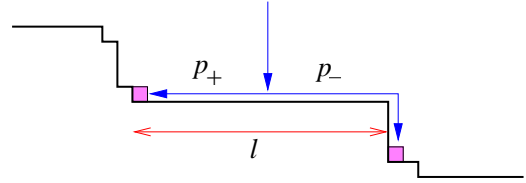}
\caption{Illustration of the model. Particles are deposited
onto the terraces and attach to the ascending (descending) step with
probability $p_+$ ($p_-$). The attachment probabilities depend on 
the terrace width $\ell$ through (\ref{eq:probabilities}).} \label{model}
\end{figure}

We consider $S$  steps of unit height located at positions $x_1$, $x_2$,
..., $x_S$ of a one-dimensional lattice with $L$ sites
and periodic boundary conditions; the slope of
the vicinal surface is $m_0 = S/L$. 
The width of the $i$-th terrace between
steps $i$ and $i-1$ is ${\ell}_i = x_i - x_{i-1}$. Due to the absence
of step-step repulsion two or more steps can be located
at the same position; correspondingly some terrace widths may be
zero. Particles are uniformly
deposited onto the sample. A particle deposited on a terrace of
width ${\ell}$ moves immediately to one of the
bordering steps and is incorporated with the probabilities
\begin{equation}
p_\pm=\frac{1}{2}\frac{1 \pm
b+d\,{\ell}}{1+d\,{\ell}}
\label{eq:probabilities}
\end{equation}
at the ascending ($p_+$) or descending ($p_-$) step,
respectively (Fig.\ref{model}). The \textit{balance} $b$ is a measure for the
attachment asymmetry, while the \textit{inverse diffusivity}
$d$ determines how the asymmetry depends on the width of the
terrace: For $d=0$ the attachment probabilities $p_\pm$ are independent
of the terrace width, while for $d > 0$ the attachment becomes
symmetric when $\ell \gg 1/d$. 
Note that in the symmetric case ($b=0$) the value of $d$ is irrelevant. 
For $d = 0$ and $b = 1$ the model is
exactly solvable, and the terrace widths have a Poisson distribution \cite{Krug95}.

The form (\ref{eq:probabilities}) follows from the solution of the
stationary diffusion equation for the adatoms on the terrace
\cite{Krug04,Politi96}, which leads to the following expressions for
the model parameters in terms of the adatom diffusion coefficient  
$D$ and the attachment rates $k_\pm$ \cite{Krug04,schwoebel}
\begin{equation}
\label{b}
b=\frac{k_+ - k_-}{k_+ + k_-}
\end{equation}
\begin{equation}
\label{d}
d= \frac{1}{D/k_+ + D/k_-}\;\; .
\end{equation}
A normal Schwoebel effect ($k_+ > k_-$) corresponds to $b > 0$, while
an iES effect implies $b < 0$. The quantity $d$ is the inverse of the sum of the
kinetic lengths $D/k_\pm$ \cite{Krug04}. 

\begin{figure}

\includegraphics[scale=0.8]{%
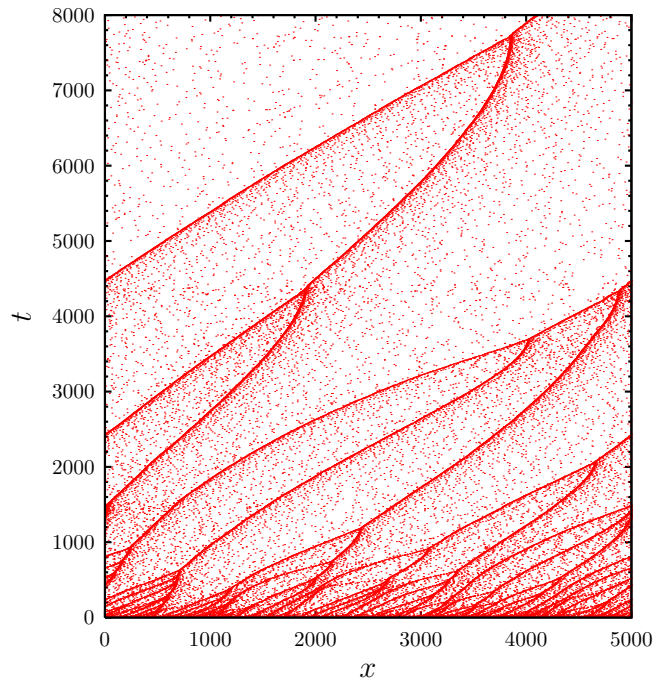}

\caption{Spatiotemporal evolution of the step configuration on a vicinal
surface of mean slope $m_0 = 0.2$. Time is increasing from bottom to top. 
The positions of the
steps are shown as dots. The sample size in the
horizontal direction  is $L = 5000$. Parameters are
$d = 0$ and $b = - 0.6$.} \label{bunches}
\end{figure}

As initial condition, the steps are placed at random.  
We use sample sizes ranging from $L = 5\times 10^4$ to
$10^6$. Time is measured in number of monolayers (ML), and simulations
were performed up to $10^{7}$ ML. As expected, 
the model exhibits step bunching when $b<0$. An example of the formation and
time evolution of bunches for $d=0$ is shown in Fig.
\ref{bunches}. 
Bunches and steps move to the right.
Eventually in the stationary
regime only a single bunch remains which moves with a 
constant velocity. In the case $d>0$ the coarsening is slower but
the space-time plot is qualitatively similar. We observed
that the bunch velocity is generally 
increasing with decreasing $d$ 
and increasing $\vert b \vert$ (see Sect.\ref{Stationary} for further
discussion).

\begin{figure}
\includegraphics[scale=0.9]{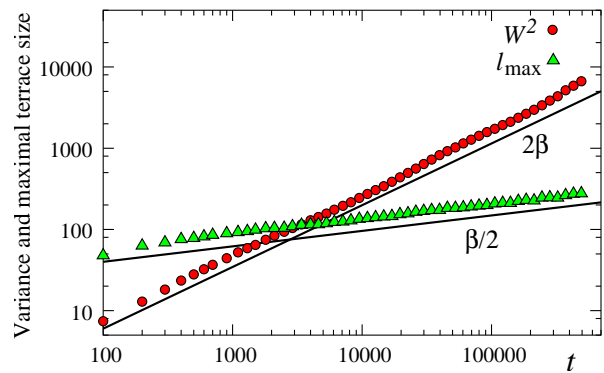}
\caption{Squared surface width $W^2$ and maximal terrace size $\ell_{\mathrm{max}}$ as 
a function of time for a system of size $L=2 \times 10^5$, slope 0.1, and
parameters $b=-1$, $d=0.1$. The bold lines have slope $2 \beta = 0.76$ and
$\beta/2 = 0.19$, respectively.}
\label{Width}
\end{figure}

The simplest way to characterize the surface morphology is to
measure the surface width $W$ relative to the mean tilted
surface \cite{Krug97b}. For $b \geq 0$ no step bunches form and
the surface is roughened only by fluctuations. For $b > 0$ the
model belongs to the Edwards-Wilkinson universality
class and $W \sim t^{1/4}$ \cite{Krug95}. For $b = 0$
we find numerically that $W \sim t^{1/3}$, which is consistent
with the idea \cite{Krug97} that the symmetric model
belongs to the conserved Kardar-Parisi-Zhang universality class \cite{Krug97b}.
In the step bunching regime $b < 0$ the surface width should asymptotically
become proportional to the bunch spacing $\cal L$. Indeed we find that
$W$ increases faster than $t^{1/3}$ when $b < 0$ (see Fig. \ref{Width}), 
but the large background contribution to $W$ 
which arises from random fluctuations makes this
quantity less suited for a precise determination of the coarsening
behavior. We therefore 
developed an alternative coarsening measure which we
describe next. 

\begin{figure}

\includegraphics[scale=0.8]{%
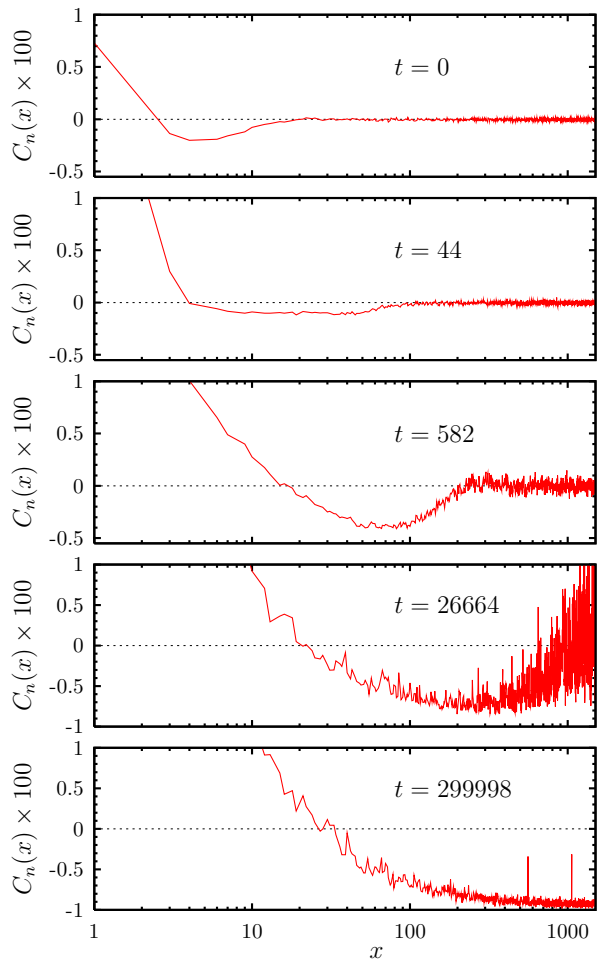}
\caption{Density-density correlation function for a system of size
$L=3000$, parameters
$b=-0.3$ and $d=0.01$, and mean slope of the vicinal surface equal to
$0.1$.
The data are averaged over 500 independent runs.
} 
\label{fig:corfun}
\end{figure}

\section{Coarsening behavior}

There are different possible ways to quantitatively analyze the
coarsening process. In analogy with phase ordering kinetics
\cite{Bray94}, we calculated
the density-density correlation function 
\begin{equation}
\label{C}
C_n(x,t)=\frac{1}{L} \sum_y \langle n(y,t)n(y+x,t)\rangle\;,
\end{equation}  
the step density being defined as 
\begin{equation}
\label{n}
n(x,t)=\sum_{i=1}^S \delta(x-x_i(t))\;,
\end{equation} 
where $\delta(x)=1$ for $x=0$ and zero otherwise. 

The typical time evolution of the correlation function is plotted in 
Fig. \ref{fig:corfun}. 
We observed a minimum in the density-density correlation function,
which shifts to larger values during growth and ultimately saturates
at half the system size. We did not observe any significant secondary
maximum in $C_n(x,t)$, indicating that there is no
systematic periodicity in the bunch configurations. 
The position of the minimum is a good indicator of the typical bunch
distance. On the other hand, it is hardly measurable with sufficient
precision. Even after averaging over $500$ independent runs the
results are not satisfactory. Therefore, we turned to 
other quantities to effectively capture the dynamics of bunching.

We may operationally define a composite step of size $k$ as an object in
which there are at least $k$ steps at the same
position. For $k=1$ we are dealing with the steps themselves, while
$k > 1$ corresponds to composite steps in the strict sense.  
Denote by $S_k$ the number of such composite steps and
$x_{k,i},i=1,2,...,S_k$ their positions. 
If we identified the step bunches with composite steps and measured
the average bunch distance as $L/S_k$, we would make
a systematic error, because  
often two such ``bunches'' come very
close one to the other, effectively making up a single larger bunch. 
We need to eliminate these short distances. To this end we define the quantity
\begin{equation}
\Delta_k =\frac{1}{L}\sum_{i=1}^{S_k}\langle (x_{k,i} - x_{k,i-1})^2\rangle ,
\end{equation}
in which  larger distances dominate over shorter ones.
Therefore, the quantity $\Delta_1$ measures the effective distance between
steps in regions of low density of steps, i.e. around the
middle between the bunches. On the other hand, $\Delta_k$ for $k>1$
measures the effective distance between bunches, irrespective of the
elementary steps dispersed between bunches. 

We compared the time evolution of $\Delta_k$ for various values of
$k$. As expected, the behavior differs qualitatively for $k=1$ and
for $k>1$. The quantity $\Delta_1$ grows with time and eventually
saturates at a value which is significantly lower than the system size. 
The saturated value increases with increasing $|b|$ and approaches the
system size for $b\to -1$. 

On the other hand, the quantities $\Delta_k$ for $k>1$ 
exhibit first a transient period with a fast decrease 
and then grow until they reach a value very close to the system size. 
 (In principle the quantity is 
 always lower than the system size, but we observed that this
 difference becomes negligible when we increase the system size. This
 fact can be easily undersood from the form of the stationary bunch
 profile, as will be shown below in Sect.\ref{HeightProfile}). 
The behavior depends on $k$ only in the initial transient period, which
is longer for larger $k$. However, the universal long time behavior,
which is the subject of this work, is found to be
independent of $k>1$. Therefore, we will concentrate on the quantity
$\Delta_2$ in the following.

\begin{figure}
\includegraphics[scale=0.8]{%
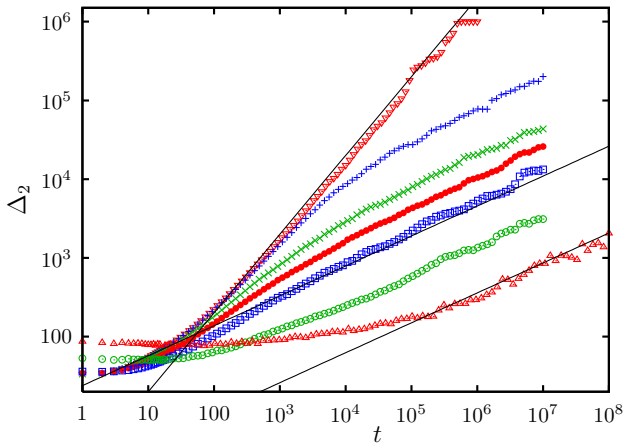}
\caption{Time evolution of the distance between bunches measured
by $\Delta_2$ for $b= - 0.9$, and varying
diffusion parameter $d = 0$, $0.0001$, $0.001$, $0.003$, $0.01$,
$0.1$, and $1$ (from top to bottom at time $10^5$ ). System sizes are
$10^6$,  $10^6$, $5\times 10^5$, $5\times 10^5$,
$10^5$, $10^5$,  and $5\times 10^4$, respectively. The lines are
powers $\propto t^1 $ and $t^{0.38}$. The mean slope of the vicinal surface
is 0.2.}

\label{fig:data}

\end{figure}

\begin{figure}

\includegraphics[scale=0.8]{%
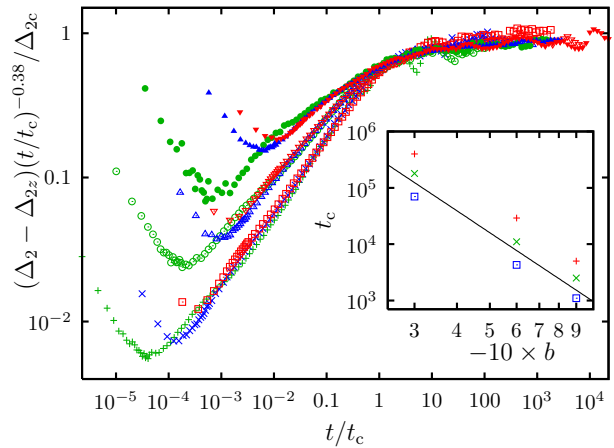}
\caption{Scaled time evolution of the effective distance between bunches
measured by $\Delta_2$ for three values of the balance and
three values of the inverse diffusivity. Groups of curves correspond to
diffusion parameters $d = 0.01$, $0.001$, $0.0001$
from top to bottom.
Each group consists of data for $b= -0.9$ ($\boxdot$, $\triangledown$,
$\blacktriangledown$), $-0.6$ ($\times$, 
$\vartriangle$, $\blacktriangle$),
and $-0.3$ ($+$, $\odot$, $\bullet$ ). The slope 
of the vicinal surface and the sample sizes are the same as in
Fig. \ref{fig:data}. 
The inset shows the values of the crossover time $t_{\rm c}$ used in the
main plot. The values of $d$ are $10^{-4}$ ($+$), $10^{-3}$
($\times$), and $0.01$ ($\boxdot$). 
The line is a power law $\propto (-b)^{-4}$.
}

\label{fig:collapse}

\end{figure}

We investigated the time evolution of the distance between bunches
in different regimes depending on the parameters $b$ and
$d$. Typical results are shown in Fig. \ref{fig:data} 
for $b= -0.9$ and several
values of the diffusion parameter $d$ (the results for other values of
$b<0$ are similar). 
For $d=0$,
linear behavior $\Delta_2
\propto t$ is reached after some transient, i.e. $\beta \approx 1$. 
On the other hand, the situation for 
nonzero $d$ is more complex. After the initial transient, there is an
intermediate behavior with approximately power-law character, but the
exponent strongly depends on the value of $d$, while the dependence on
$b$ was very weak.

Eventually, after some crossover time $\simeq t_{\mathrm c}$ a stationary
power-law regime is 
reached, which in our data extends over several orders of magnitude in
time. The effective exponent $\beta$ is only
weakly varying with the nonzero $d$ and assumes the
asymptotic value $\beta\simeq 0.38$. We conjecture that this regime is
universal and the remaining variations in the value of the exponent
are solely due to finite size effects.

In order to confirm that the asymptotic behavior is universal, one
can rescale the raw data for different nonzero $d$, plotting them as
functions of  $t/t_{\mathrm c}$ in the form
\begin{equation}
\label{scalingform}
\Delta_2(t;b,d)=\Delta_{2z}+
(t/t_{\mathrm c})^{0.38}\,\Delta_{2\mathrm c}\,F(t/t_{\mathrm c}).
\end{equation} 
The results for three values of $d$ and
three values of $b$ are shown in Fig. \ref{fig:collapse}. 
Neglecting the initial transient, we see that the
scaling funtion $F(x)$ approaches unity for $x\gg 1$, marking the
asymptotic universal regime, and behaves close
to a power-law $F(x)\sim x^{\tau}$ for $x\ll 1$, 
which characterizes the intermediate regime;
the exponent $\tau$ depends on $d$ but is independent of $b$. 
The initial offset $\Delta_{2z}$ was introduced for convenience.

The crossover time $t_{\mathrm c}$ between
the intermediate and asymptotic regimes turned out to
be rather large ($t_{\mathrm c} \simeq 10^4$ for $d=0.001$ and $b=-0.9$) 
and it increases rapidly with decreasing $|b|$, as can be seen in the inset of
Fig. \ref{fig:collapse}, where behavior consistent with a power law
$t_{\mathrm c}\sim (-b)^{-4}$ is apparent. The dependence on $d$ turns out to
be non-monotonic: The crossover time decreases 
up to about $d=0.01$ and then increases again.
This shows that different crossover mechanisms are acting for small
and large $d$, respectively. 

Indeed, for very small $d$ the dependence of the 
attachment asymmetry on the terrace size is felt only once the terraces
become sufficiently large, and the system behaves initially as if
$d=0$ and $\beta = 1$; correspondingly the coarsening exponent approaches
its asymptotic value from above in this case. Conversely, for large $d$
the asymmetry in the attachment probabilities (\ref{eq:probabilities})
is always very  
small; consequently it takes a long time for the bunches to form, and 
the asymptotic coarsening exponent is approached from below. 


The asymptotically linear behavior of the bunch size for $d=0$, $\beta
\approx 1$, is well confirmed by measurements of the surface width
$W$, and the data for $d=0.1$ shown in  Fig. \ref{Width} are
consistent with the estimate $\beta \approx 0.38$ obtained from
$\Delta_2$. In addition, Fig.\ref{Width} displays the evolution of the
maximal terrace width $\ell_{\mathrm{max}}$ in the system. This length
scale is seen to scale with a distinct exponent which is approximately
given by $\beta/2 \approx 0.19$. We will return to the behavior of 
$\ell_{\mathrm{max}}$ below in Sect.\ref{HeightProfile}. 

\section{Stationary bunches}

\begin{figure}
\includegraphics[scale=0.7]{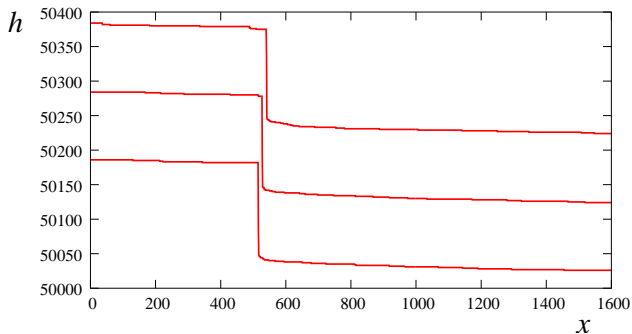}
\caption{Height profiles of a stationary bunch for $d=0$, $b=-0.1$. The figure
shows a surface with $S = 160$, $L=1600$ after deposition of 
50000, 50100 and 50200 ML, respectively.}
\label{Profiles}
\end{figure}

The final state of the system for any $b<0$ is found to be a single bunch
propagating at constant velocity (Fig.\ref{Profiles}). 
The analysis of these stationary
configurations provides important clues to the different coarsening
behaviors observed for $d=0$ and $d>0$, respectively.

\label{Stationary}

\subsection{Bunch speed}

\begin{figure}
\includegraphics[scale=0.8]{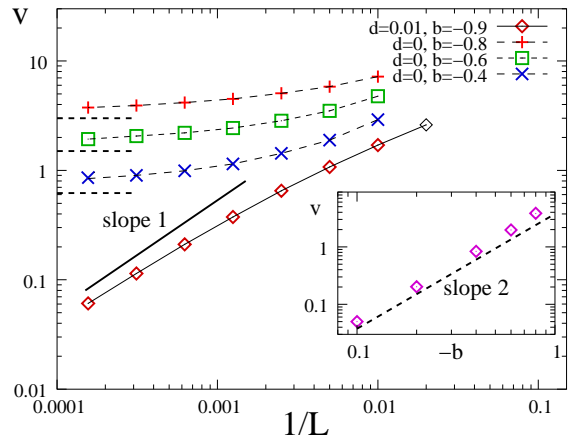}
\caption{Propagation speed $v$ of stationary bunches. The main plot
shows the dependence 
of $v$ on the system size $L$, contrasting the behavior for $d=0$ and $d > 0$.
The inset shows the bunch speed for $d = 0$, extrapolated to $L =
\infty$, as a function 
of $\vert b \vert$; the extrapolation is based on data up to $L=1600$
and assumes 
a leading correction of order $1/L$. All data
were obtained by time averages in the stationary regime. The surface slope was
$0.1$ for $d=0$ and $0.2$ for $d=0.01$. The full bold line illustrates the
behavior $v \sim 1/L$. The dashed bold lines are the \emph{upper} bounds
on $v$ derived in Sect.\ref{Theory}. The bounds in the main figure
were derived from 
the algebraic equation (\ref{alg}), while the bound in the inset is the 
inequality (\ref{boundv}) which becomes exact for $\vert b \vert \to 0$.}
\label{Speed}
\end{figure}

Figure \ref{Speed} shows numerical results for the stationary bunch speed
$v$. For $d=0$,
$v$ approaches a finite limiting value $v_\infty$ with increasing
system size; this is also evident from the space-time
plot in Fig.\ref{bunches}. The bunch speed is easily determined for the limiting
case $d=0$, $b=-1$. In this limit single steps cannot
detach from the bunches, and the formation of composite steps is irreversible.
The motion of a bunch is solely due to the atoms deposited onto
the trailing terrace behind the bunch, which has length $L$ in the stationary
state. To move the bunch containing all $S$ steps laterally
by one lattice spacing, $S$ atoms have to be deposited on this terrace;
the bunch speed is therefore
$v_\infty = L/S = 1/m_0$.
The limiting bunch speed decreases with decreasing $\vert b \vert$ and vanishes
as $v_\infty \sim b^2$ for $b \to 0$ (see inset of Fig.\ref{Speed}). 
An analytic explanation of this behavior
will be given below in Sect.\ref{Theory}.

For $d > 0$ the bunch speed decreases indefinitely with increasing $L$. The simulation
results shown
in Fig.\ref{Speed} are consistent with a behavior 
\begin{equation}
\label{nu}
v(L) \sim L^{-\nu}
\end{equation}
with $\nu \approx 1$, but owing to the significant curvature of the 
data, the asymptotic value of $\nu$ cannot be accurately determined. 

Following a simple scaling argument due to Chernov \cite{Chernov61},
we can try to relate
the exponent $\nu$ in (\ref{nu}) to the coarsening exponent
$\beta$. At a time when the typical bunch spacing is $\cal{L}$, the bunch
speeds are of order ${\cal{L}}^{-\nu}$. Assuming that there is only a single
scale in the problem, the velocity difference $\Delta v$ between two
bunches is of the  
same order. The time $t^\ast$ until the coalescence of two
bunches can then be
estimated as $t^\ast \sim {\cal{L}}/\Delta v \sim {\cal{L}}^{1
+ \nu}$, and 
reversing this relationship one obtains the expression
\begin{equation}
\label{betanu}
\beta = \frac{1}{1 + \nu}
\end{equation}
for the coarsening exponent.  
For $d=0$ we have seen that the bunch speed remains finite for $L \to \infty$, hence
$\nu = 0$ and $\beta = 1$ in agreement with the coarsening simulations. 

For $d > 0$
the value $\nu \approx 1$ implies $\beta \approx 1/2$, which is larger than our
numerical estimate $\beta \approx 0.38$. Although the
data for $v(L)$ in Fig.\ref{Speed} are probably not asymptotic,
it seems unlikely that they will ever reach the large slope $\nu \approx 1.6$ required
to reproduce our value for $\beta$. In fact, our simulations show clearly
that, besides the coalescence of bunches, the 
exchange of steps between bunches of different size plays an important
role in the coarsening process. In addition, the existence of the 
distinct length scale $\ell_{\mathrm{max}}$ (see Fig.\ref{Width}) invalidates
the assumption that the bunch spacing is the only length scale in the problem.
Thus, while Chernov's relation (\ref{betanu}) helps to connect
the different coarsening behaviors for $d=0$ and $d>0$ with the difference
in the size dependence of the bunch speed on a qualitative level, it is
not quantitatively satisfied for our model.

\begin{figure}

\includegraphics[scale=0.8]{%
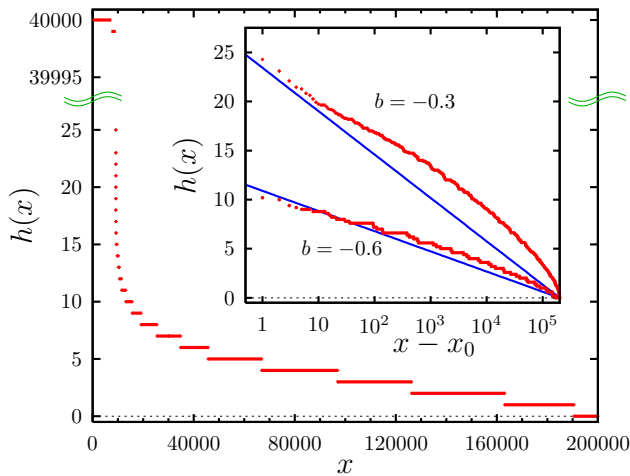}
\caption{Stationary bunch profile at time $1\times 10^{7}$  for
sample size $L = 2 \times 10^5$ with slope 0.2  for $d = 0$ and
 balance $b =-0.3$. Inset shows log-linear
plot of profile relative to the center of bunch $x_0$, averaged
over 11 (for $b=-0.3$) and 5 (for $b=-0.6$) independent runs. The
lines are predictions from the formulae (\ref{hprofile}) and (\ref{alg}).}

\label{fig:profile} 

\end{figure}

\subsection{Height profile}
\label{HeightProfile}

We measured the height profile between bunches in the stationary regime. We show in Fig.
\ref{fig:profile} the results for $d=0$ and two values $b=-0.3$ and
$b=-0.6$. Qualitatively we observe that for larger values of $-b$ the
bunch becomes steeper in its very center, while the rest of the bunch,
farther from the center, is more gradual. 
Since there is no repulsion between steps they
tend to accumulate in the center of bunch (position denoted $x_0$) and
to the right of it, while the region on the left-hand side is much
more flat. This means that the bunch shape is strongly asymmetric. 
Due to the discreteness of
the lattice it it very difficult to infer precisely the singular behavior near
the center of bunch, but the averaged data in the inset show 
that the profile far from the
center can be described by a logarithmic behavior of the form
\begin{equation}
\label{hprofile}
h(x_0) - h(x) \approx A \ln(x - x_0).
\end{equation}
The logarithmic singularity observed in the bunch shape is
very weak and in all cases investigated most of the mass of the bunch
is concentrated on two sites in its center. Indeed, setting 
$x - x_0 = L$ in (\ref{hprofile}) it is seen that the number $S_t$
of \textit{terrace steps} which are not contained in the center of the bunch
grows only logarithmically with $L$. 
This is the ultimate reason why the effective quantity $\Delta_2$
measures quite well the distance between bunches.

\begin{figure}

\includegraphics[scale=0.9]{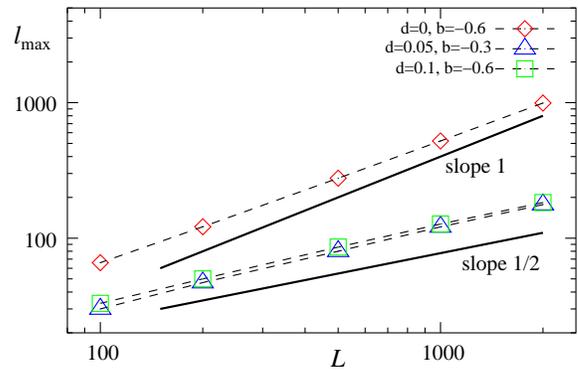}
\caption{Maximal terrace size $\ell_{\mathrm{max}}$ for stationary bunches as
a function of system size $L$. The mean slope is $m_0 = 0.2$.
The data were obtained as time averages over single runs.}
\label{fig:lmax} 

\end{figure}

The logarithmic height profile implies that the surface slope $m(x) = \partial h/\partial x$
decays as $m(x) \approx - A/x$ far away from the bunch. The size $\ell_{\mathrm{max}}$
of the largest terrace in the system, which is found at a distance of order $L$
from the bunch, is then of order $\ell_{\mathrm{max}} \sim - m(L) \sim L$.
This is consistent with measurements of $l_{\mathrm{max}}$ for stationary
bunches with $d=0$ (Fig.\ref{fig:lmax}). On the other hand, for $d > 0$ the largest terrace is much
smaller, and scales approximately as $\ell_{\mathrm{max}} \sim L^{1/2}$. This
matches the time-dependent behavior of $\ell_{\mathrm{max}}$ shown
in Fig.\ref{Width}.

\subsection{Analytic description of stationary bunches}
\label{Theory}

In this section we provide an analytic derivation of several of the 
numerically observed properties of stationary bunches for the case $d=0$
\cite{Comment:Slanina}. 
A convenient starting point is the continuum evolution
equation 
\begin{equation}
\label{continuum}
\frac{\partial h}{\partial t} + \frac{\partial}{\partial x} 
\left[- \frac{\vert b \vert}{2 m} + \frac{1}{6 m^3} \frac{\partial m}{\partial x} 
\right] = 1,
\end{equation} 
which has been obtained for the deterministic version of the 
present model through an essentially rigorous coarse graining
procedure \cite{Politi96,Krug97}. Here 
$m = \partial h/\partial x$ is the surface slope, the time scale
is chosen such that the deposition flux equals unity, and $b$ is assumed
negative. A continuum description of the type (\ref{continuum}) is expected
to be valid far ahead of the bunch, where the step spacing is large compared
to the atomic scale and lattice effects are negligible.  

We are looking for solutions of (\ref{continuum}) which describe 
a bunch of height $S$ (in units of the vertical lattice spacing)
in a system of length $L$, moving laterally at speed $v$. This implies the ansatz
\begin{equation}
\label{moving}
h(x,t) = f(x - vt) + \Omega t,
\end{equation}
where the last term accounts for the fact that also the regions between
bunches grow vertically due to the terrace steps that move across these
regions \cite{Stoyanov04} (see Fig.\ref{Profiles}). The constant unit flux on the right hand side of 
(\ref{continuum}) implies the sum rule \cite{Popkov04}
\begin{equation}
\label{sumrule}
\Omega + \frac{S}{L} v = 1
\end{equation}
connecting the vertical and lateral growth rates.
Inserting (\ref{moving}) into (\ref{continuum}) one obtains the ordinary
differential equation 
\begin{equation}
\label{ODE}
-v g + \frac{d}{d\xi} \left[- \frac{\vert b \vert}{2 g} + \frac{1}{6 g^3} \frac{dg}{d\xi} 
\right] = 1 - \Omega
\end{equation}
where $\xi = x - vt$ and $g = df/d\xi$ is the surface slope in the comoving
frame. The
first term on the left hand side of (\ref{ODE}) can be neglected relative
to the right hand side when $g \ll m_0$, which is true far ahead of the bunch. Then it
is readily verified that the equation admits a solution of the 
form $g(\xi) = - A/\xi$, consistent with the height profile (\ref{hprofile}).
The coefficient $A$ satisfies the quadratic
equation 
\begin{equation}
\label{quadratic}
(1 - \Omega)A^2 - A \vert b \vert/2 + 1/6 = 0.
\end{equation} 
Using (\ref{sumrule}),
one finds that (\ref{quadratic}) 
has real solutions only when the bunch speed
satisfies the inequality
\begin{equation}
\label{boundv}
v \leq \frac{3}{8} \frac{b^2}{m_0}.
\end{equation}
Whereas the proportionality $v \sim b^2$ suggested by this relation
is well confirmed by the data in the inset of 
Fig.\ref{Speed}, the bound on the numerical coefficient
seems to be weakly violated by the data. However, since the convergence
of $v$ with increasing $L$ is quite slow, we believe that our simulations
can still be regarded to be consistent with (\ref{boundv}). Assuming that
(\ref{boundv}) is satisfied essentially as an equality, we find that
the coefficient $A$ is given by 
\begin{equation}
\label{A}
A = \frac{2}{3 \vert b \vert}.
\end{equation}

The relations (\ref{boundv}) and (\ref{A}) can be improved by analyzing
the discrete evolution equations for the mean step positions
$X_i = \langle x_i \rangle$. For $d=0$ they take the simple linear form
\cite{schwoebel,Krug97}
\begin{equation}
\label{linear}
\frac{d X_i}{dt} = \frac{1}{2} (1 + b) (X_{i+1} - X_i) + 
\frac{1}{2} (1 - b) (X_{i}  - X_{i-1}).
\end{equation}
The moving bunch is described by a solution of the form
\begin{equation}
\label{discrete}
X_i (t) = \phi(i + \Omega t)+ vt,
\end{equation}
where $\Omega$ and $v$ have the same meaning as in (\ref{moving}).
Inserting (\ref{discrete}) into (\ref{linear}) one obtains the difference-differential
equation 
\begin{equation}
\label{phi}
v + \Omega \frac{d \phi}{d \zeta} = 
- b \phi(\zeta) + \frac{1}{2} (1 + b) \phi(\zeta +1) - \frac{1}{2} (1 - b) \phi(\zeta - 1)
\end{equation}
for the function $\phi(\zeta)$, where $\zeta = i + \Omega t$. The expected logarithmic
height profile (\ref{hprofile}) corresponds to an exponential increase of 
$\phi$ as $\phi(\zeta) \sim e^{\zeta/A}$. For large $\zeta$ the constant $v$  on the
left hand side of (\ref{phi}) can be neglected. Inserting the 
exponential ansatz into (\ref{phi}) yields the algebraic equation
\begin{equation}
\label{alg}
\Omega = A \{ \sinh(1/A) + \vert b \vert [1 - \cosh(1/A)] \},
\end{equation}
which reduces to (\ref{quadratic}) for small $\vert b \vert$ and large $A$.
For general values of $\vert b \vert$, (\ref{alg}) has real solutions only when
the bunch speed $v$ is smaller than an upper bound $v_{\mathrm{max}}(b)$, which
reduces to (\ref{boundv}) for small $\vert b \vert$. In Figs.\ref{Speed} and
\ref{fig:profile} we compare the predictions for $v_{\mathrm{max}}$ and $A$ derived from 
(\ref{alg}) with our numerical results. For $A$ we use the value obtained from
(\ref{alg}) under the assumption that $v = v_{\mathrm{max}}$, which leads to 
a rather satisfactory agreement.


\section{Conclusions}

In this paper we have developed and studied a one-dimensional
stochastic growth model which contains some essential features
of the formation and coarsening of step bunches. 
Despite its simplicity, the model displays two distinct
scaling regimes. In the limit of fast terrace diffusion 
($d=0$) we have numerically established a linear coarsening 
law ($\beta = 1$) and related this behavior to the fact that
the propagation speed of bunches tends to a finite value
with increasing bunch size. 

When the diffusivity is finite
($d > 0$) the attachment asymmetry decreases with increasing
terrace size and the coarsening is slowed down. 
After having accounted for a complicated crossover behavior, 
we identified the asymptotic value of the coarsening exponent,
which we estimate to be $\beta \simeq 0.38$. 
This exponent is different from exponents found in 
previous studies of step bunching, and it does not seem to 
follow from any simple scaling argument. It is robust, in the sense
of being independent of the precise values of the model parameters,
provided $b < 0$ and $d > 0$. 

Similar to earlier studies of kinetic roughening
\cite{smilauer94}, our work shows that
finding the true asymptotic exponent may require a considerable amount of
computer power due to a long
intermediate regime with non-universal, apparent power-law behavior. 
The situation becomes even more difficult when we try to observe the
crossover from $\beta=1$ to $\beta\simeq 0.38$, which
should occur for sufficiently small but non-zero $d$ and large enough
system size $L$. Currently this is beyond our capacity.  

Due to the absence of step-step interactions, our model cannot provide
an accurate description of the structure and dynamics of real step
bunches. Nevertheless our study clarifies some conceptual issues that
arise also for more realistic models. For example, although the internal width
of the bunches has been eliminated by allowing steps to coalesce,
the model still displays (for $d > 0$) a second length scale,
the maximum terrace size $\ell_{\mathrm{max}}$, which grows with 
an exponent that is distinct from the coarsening exponent. This invalidates
simple scaling arguments \cite{pimpinelli02,Krug04a,Chernov61} which assume the bunch spacing
to be the only scale in the problem, and which would imply a ''superuniveral''
coarsening exponent $\beta = 1/2$ \cite{Liu98}. 

A second important advance
of our work is the demonstration that continuum evolution equations can 
be used to quantitatively describe the profile and speed of moving 
step bunches. So far the use of continuum approaches in step bunching
has been limited to scaling arguments \cite{pimpinelli02} and 
to the study of stationary bunch configurations \cite{Liu98,Krug04b}.  

To conclude, we speculate that the relationship between growing stepped surfaces
and interacting particle systems (specifically, zero range processes 
\cite{Evans00}) that was pointed out some time ago \cite{Krug97} may be 
usefully exploited to gain further insight into the coarsening behavior
of step bunches. In the language of interacting particle systems, step bunching
is an example of a \textit{condensation transition}, where a macroscopic 
fraction of the particles (i.e., the single height steps in our model)
condenses into a few or a single site. The dynamics of such condensation
transitions is of great current interest \cite{Evans00,Grosskinsky03}, and
closely related concepts appear e.g. in granular physics \cite{Hellen02,vanderMeer04}
and traffic research \cite{Krug97,Chowdhury00}.    

\begin{acknowledgments}

This work was supported by  the Grant Agency of the Czech Republic
(Grant No. 202/03/0551), by Volkswagenstiftung, and by DFG within
SFB 616 \emph{Energiedissipation an Oberfl\"achen}. J.K. acknowledges
the contribution of Claudio Castellano in the early stages of this
work, and wishes to thank Stoyan Stoyanov and Vladislav Popkov for
useful interactions.  

\end{acknowledgments}

\end{document}